\documentclass
[preprint,showpacs,byrevtex,10pt,twocolumn,prl,letterpaper,tightenlines]{revtex4}%
\usepackage{amsfonts}
\usepackage{amsmath}
\usepackage{amssymb}
\usepackage[dvips]{graphicx}
\usepackage{hyphenat}%
\providecommand{\U}[1]{\protect\rule{.1in}{.1in}}

\begin{document}
\preprint{ }
\title{Infrared-Induced Sluggish Dynamics in the GeSbTe Electron Glass.}
\author{Z. Ovadyahu}
\affiliation{Racah Institute of Physics, The Hebrew University, Jerusalem 91904, Israel }

\pacs{72.20.-i 72.80.Ng 72.40.+w 78.47.da}

\begin{abstract}
The electron-glass dynamics of Anderson-localized GeSbTe films is dramatically
slowed-down following a brief infrared illumination that increases the system
carrier-concentration (and thus its conductance). These results demonstrate
that the dynamics exhibited by electron-glasses is more sensitive to
carrier-concentration than to disorder. In turn, this seems to imply that
many-body effects such as the Orthogonality Catastrophe must play a role in
the sluggish dynamics observed in the intrinsic electron-glasses.

\end{abstract}
\maketitle

The interplay between disorder and Coulomb interactions makes Anderson
insulators a natural candidate to exhibit out-of-equilibrium transport
properties characteristic of glasses \cite{1,2,3,4,5,6,7,8,9,10,11,12}. Glassy
features such as slow relaxation, ageing, and other memory effects related to
such an electron-glass (EG) scenario have been observed in a number of systems
\cite{13}. A feature common to all these systems is relatively high
carrier-concentration n. Intrinsic electron-glass effects with relaxation
times longer than seconds seem to be peculiar to Anderson insulators with n
spanning the range of $\approx$4x10$^{\text{19}}$cm$^{\text{-3}}$ (crystalline
indium-oxide) to $\simeq$10$^{\text{22}}$cm$^{\text{-3}}$ (beryllium)
\cite{13}. The relaxation time $\tau$ in amorphous indium-oxide with n%
$<$%
8x10$^{\text{19}}$cm$^{\text{-3}}$ dropped by more than two orders of
magnitude when n was reduced by a mere factor of $\approx$three \cite{14}.
This is also consistent with the absence of intrinsic EG effects in
lightly-doped semiconductors; experiments on phosphorous-doped silicon
exhibited relaxation times much shorter than a second \cite{15}.
Condensed-matter phenomena that favor high carrier-concentration is usually a
many-body effect (screening, superconductivity). It is therefore of interest
to study this aspect in a controlled manner. The fast dependence of $\tau$ on
n observed in \cite{14} employed amorphous indium-oxide samples with different
\textit{compositions} to obtain different carrier-concentrations. The recent
observation of coexistence of electron-glass phase with
persistent-photoconductivity in GeSb$_{\text{x}}$Te$_{\text{y}}$ \cite{16}
gives us a unique opportunity to check this issue in a \textit{single} sample.
Persistent-photoconductivity (PPC) is used in this work to change the
carrier-concentration of the system by a considerable amount without changing
the main structural features of the sample \cite{17}. It is shown that, in the
PPC-state, the electron-glass dynamics is slowed down by a more than an order
of magnitude relative to the dark-state. This dramatic effect may help in
elucidating the mechanisms responsible for the slow dynamics exhibited by
intrinsic electron-glasses as it hints on the importance of many-body effects.

Samples used for this study were prepared by e-gun depositing a
GeSb$_{\text{2}}$Te$_{\text{5}}$ alloy onto room temperature substrates. The
substrates were 0.5$\mu$m SiO$_{\text{2}}$ layer thermally grown on
$<$%
100%
$>$
silicon wafers having bulk resistivity $\rho\simeq$ 2x10$^{\text{-3}}\Omega
$cm, deep into the degenerate regime. The wafers were employed as the gate
electrode in\ the field-effect measurements. Polycrystalline samples of
GeSb$_{\text{x}}$Te$_{\text{y}}$ were obtained by crystallizing the
as-deposited (amorphous) films$\ $a~temperature of 460$\pm$5K. Conductivity of
the samples was measured using a two terminal ac technique employing a
1211-ITHACO current preamplifier and a PAR-124A lock-in amplifier.
Measurements were performed with the samples immersed in liquid helium at
T$\approx$4.1K held by a 100 liters storage-dewar. This allowed up to two
months measurements on a given sample while keeping it cold (and in the dark).
Optical excitations employed AlGaAs diode operating at $\approx$0.88$\pm
$0.05$\mu$m, mounted on the sample-stage~at a distance of $\approx$12mm from
the sample. The diode was energized by a computer-controlled Keithley 220
current-source. Fuller details of sample preparation and characterization are
described elsewhere \cite{16}.

The protocol used for assessing the electron-glass dynamics is the
two-dip-experiment (TDE). This is illustrated in Fig.1 showing a series of
conductance vs. gate-voltage G(V$_{\text{g}}$) traces taken at different times
for a GeSb$_{\text{x}}$Te$_{\text{y}}$ sample in the dark-state. The first
G(V$_{\text{g}}$) trace was taken 24 hours after the sample was cooled down to
T=4.1K and allowed to equilibrate while V$_{\text{g}}$=5.8V was held between
the sample and the gate. The resulting field-effect trace is composed of two
components; an anti-symmetric component reflecting the underlying
(thermodynamic) density-of-states (DOS), and a superimposed dip, centered
around the gate-voltage where the systems was allowed to relax (V$_{\text{g}}%
$=5.8V in this particular case). The latter, so called memory-dip (MD), is the
distinguishing feature of the intrinsic electron-glass \cite{17}, reflecting
an underlying Coulomb-gap \cite{5,9,12,13}. The width of the memory-dip is
comparable with that of the amorphous version of indium-oxide \cite{14} with
the same carrier-concentration (6$\pm$3)x10$^{\text{20}}$cm$^{\text{-3}}$. A
second trace shown in Fig.1a was taken after the gate-voltage was moved to
V$_{\text{g}}$=-5.8V and left there for 10 minutes before a new G(V$_{\text{g}%
}$) trace was taken. This trace shows a dip centered at the newly imposed
gate-voltage while the old dip is diminished in magnitude (marked in the
figure as $\delta$G$_{\text{MD}}$). This protocol is repeated to produce the
other curves shown in Fig.1a while V$_{\text{g}}$ is parked at -5.8V between
gate-sweeps taken at later times.

The inset to Fig.1 shows the time dependence of the magnitude of the old and
new MD's which, in agreement with previously studied electron-glasses, exhibit
logarithmic dependence \cite{10,18}. A typical time characterizing the
dynamics of the TDE protocol is the time where the magnitude of the old MD
equals that of the new one. This turns out to be $\approx$33~minutes, which
again is similar to the result obtained for In$_{\text{x}}$O film with
comparable carrier-concentration measured under identical protocol \cite{14}
(including history).%
\begin{figure}[ptb]%
\centering
\includegraphics[
height=3.2984in,
width=3.34in
]%
{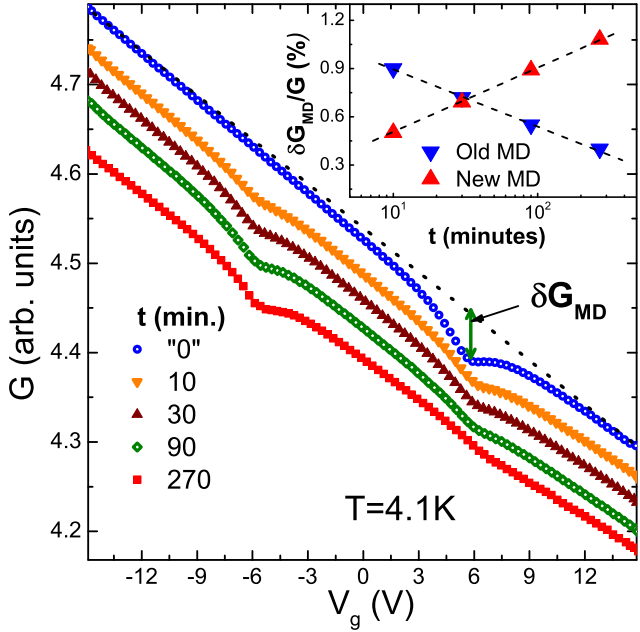}%
\caption{The two-dip-experiment performed on a GeSb$_{\text{x}}$Te$_{\text{y}%
}$ film with R$_{\square}$=4.4M$\Omega$ in the dark-state. Curves are shifted
for clarity. The dotted line depicts the underlying thermodynamic
density-of-states. Inset shows the time dependence of the old and new memory
dips (see text for details). Sample lateral dimensions are W=1~mm,
L=${\frac12}$~mm.}%
\end{figure}

This dynamics is observed when the sample is kept in the dark. A dramatic
change is observed when the system is cast into its
persistent-photoconductivity (PPC) state \cite{16}. This is achieved by
exposing the sample to the infrared source causing the conductance G to
increase by $\approx$50\%. The effect of the infrared excitation and the
ensuing conductance relaxation are shown in Fig.2
\begin{figure}[ptb]%
\centering
\includegraphics[
height=2.6193in,
width=3.34in
]%
{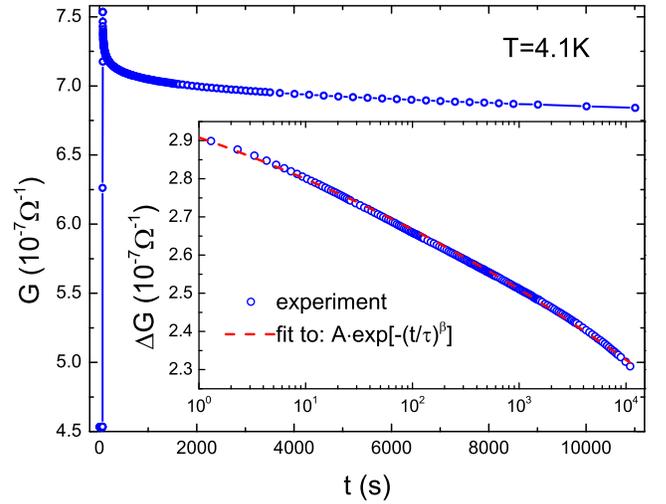}%
\caption{The time dependence of the conductance before, during, and after
exposing the sample (as in Fi1.1) to infrared (3~seconds at 1mA). The inset
depicts fitting the relaxation of G in the induced PPC-state. Fitting
parameters are: A=3.375, $\tau$=2x10$^{\text{8}}\sec$, $\beta$=0.1. }%
\end{figure}

The inset to Fig.2 depicts the relaxation of the excess conductance $\Delta$G
created by the IR excitation. This $\Delta$G(t) fits well a stretched
exponent, $\Delta$G(t)$\propto$exp[-(t/$\tau$)$^{\beta}$] with $\beta$%
=0.1$\pm$0.01. The same relaxation law with the \textit{same} $\beta$ (for
samples measured at T=4.1K) has been seen in more than dozen other
GeSb$_{\text{x}}$Te$_{\text{y}}$ films with sheet resistances R$_{\square}$
ranging from 2.5k$\Omega$ to 25M$\Omega$ \cite{16}. The infrared excitation
has also increased the magnitude of the MD as can be seen in the top curve for
G(V$_{\text{g}}$) in Fig.3.%
\begin{figure}[ptb]%
\centering
\includegraphics[
height=3.2661in,
width=3.34in
]%
{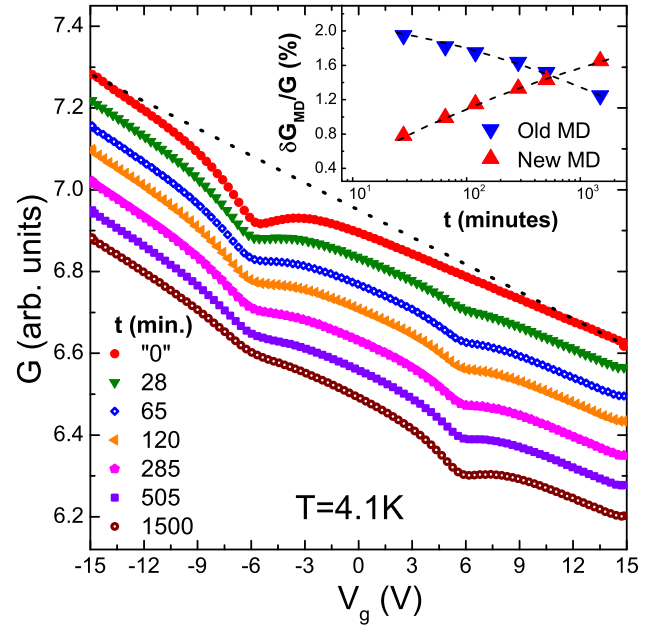}%
\caption{The two-dip-experiment as in Fig.1 but after the sample is put in its
PPC-state. Curves are shifted for clarity. The dashed lines are guides for the
eye. }%
\end{figure}
This trace was taken $\approx$22 hours after the infrared source was turned
off. It is the starting stage for a TDE protocol of the system in its
PPC-state. The same protocol was used in the G(V$_{\text{g}}$) series shown in
Fig.3 as in the previous two-dip experiment of Fig.1 but it had now to be
extended over much longer time to accommodate data points where $\delta
$G$_{\text{MD}}$(t)/G of the new MD exceeds the amplitude of the old MD. As
may be seen from the inset to Fig.3 the typical time of the TDE in this case
is $\approx$580 minutes, about 17-times longer than for the system in its
dark-state. Note (inset to Fig.3) that $\delta$G$_{\text{MD}}$(t)/G curves for
both the new and old MD's deviate from the logarithmic law that characterize
their time dependence in the dark-state TDE series (Fig.2). This presumably is
a result of the system drifting towards its dark-state with its associated
smaller MD amplitude \cite{16}. It is not however the reason for the perceived
slowdown of the dynamics in the PPC-state; the same typical time for the TDE
would be gathered by using the extrapolated crossing point for logarithmic
curves based on the first 2-3 points in the series.

The more than an order of magnitude increase in the TDE typical time (Fig.3
vs. Fig.1) was also observed on two other GeSb$_{\text{x}}$Te$_{\text{y}}$
films with R$_{\square}$=2.1M$\Omega$ and R$_{\square}$=8.5M$\Omega$. We also
confirmed that both, the enhanced magnitude of the MD and the slower dynamics
of the PPC-state are reversible; by keeping the sample for a minute or two
above \ 30-40K then re-cooling to T=4.1K the dark-state on all its previous
features are restored, including the faster dynamics.

It is difficult to get a direct experimental reading of the change in the
previously established MD in the dark-state immediately\textit{ }after turning
off the infrared source; the fast change of G(t) in the initial stage of the
PPC relaxation would overwhelm a G(V$_{\text{g}}$) measurement. There is
however an indirect way to infer something about this issue from
G(V$_{\text{g}}$) measurements performed in later times, starting from
$\approx$12~minutes after the infrared illumination. Results of these
measurements are shown in Fig.4.%
\begin{figure}[ptb]%
\centering
\includegraphics[
height=3.34in,
width=3.34in
]%
{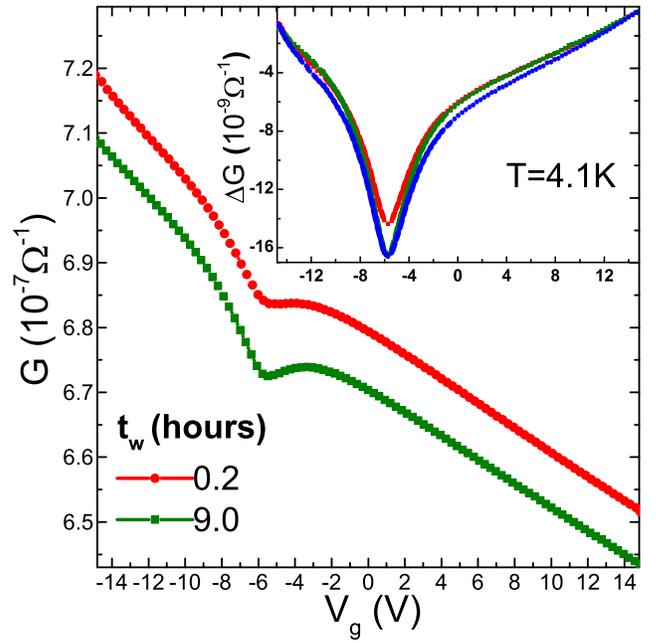}%
\caption{Conductance vs. gate-voltage sweeps for the sample taken at the
indicated times after the sample was exposed to infrared radiation (using the
protocol described in Fig.2). The inset shows the corresponding MD obtained by
subtracting a straight line from the respective G(V$_{\text{g}}$) trace (with
corresponding labels as in the main figure). The blue circles are data points
for the 12-minutes MD multiplied by a constant (1.157) to match the $\Delta$G
value of the 9 hours trace at equilibrium V$_{\text{g}}$. This illustrates
that the 12-minutes MD is wider than the long-time value \{these MD's appear
to coincide at the end-points of the V$_{\text{g}}$ measured-range
"by-construction", due to the way they were obtained from the composite
G(V$_{\text{g}}$)\}.}%
\end{figure}
Note that 12 minutes after the infrared exposure the MD amplitude $\delta
$G$_{\text{MD}}$(t)/G is $\approx$85\% of its maximum value (obtained at
intermediate times; at long times $\delta$G$_{\text{MD}}$(t)/G
actually\textit{ goes down }as the PPC-state dissipates and the dark-state
value is recovered \cite{16}). At the same time however, the \textit{shape} of
the MD differs from its asymptotic form. This can be seen in the inset to
Fig.4; normalizing the amplitude of the 12 minutes MD to that of the 9-hours
MD fails to show the simple scaling that characterizes how the memory-dip
evolves with time under other protocols. In particular, the MD shape is time
independent following quench-cooling the sample from high temperatures
\cite{19}. The wider MD shape of the 12-minutes trace is consistent with a
"hotter" or more energetic electronic state than in either the dark-state or
the asymptotic PPC-state \cite{19}.

The "hotter" nature of the 12-minutes MD suggests that the infrared exposure
disturbed the relaxed configuration of the dark-state. Judging by the $\Delta
$G produced by the illumination, the added carriers density in the PPC-state
$\Delta$n is a sizeable fraction of the carrier-concentration of the
dark-state \cite{16}, in this particular case of the order of $\approx
$10$^{\text{19}}$cm$^{\text{-3}}$. Changing the charge-density of the
interacting system naturally takes it out of equilibrium. It is less clear
that it could also slow-down its dynamics. This needs elaboration.

Relaxation from an excited state of the electron-glass proceeds by transitions
between localized states. So does the system conductance. One should be
careful however in drawing conclusions about relaxation from observations
pertinent to conductance. In the first place, there are many more sites
involve in relaxation than those participating in dc transport \cite{13,21}.
Conductance is controlled by the relatively fast transitions in the
current-carrying network \cite{20} while relaxation must also involve
transitions in the `dead-wood' regions of the system. Local dynamics in these
regions, which occupy most of the system volume, is also orders of magnitude
slower than that of the sites that are part of the current-carrying-network.
The global relaxation rate is a hierarchical process composed of
avalanche-like events propagating through the \textit{entire} system
\cite{21}. Secondly, while conductance is essentially controlled by
single-particle transitions, relaxation requires energy-diminishing events and
these ultimately necessitate simultaneous \textit{multi}-particle transitions
\cite{21}. These transitions are required to approach the lowest energy
configuration and it is plausible that their relative importance grows with
the carrier-density \cite{22,23}.

Conductance and relaxation may also differ \textit{qualitatively: }In several
electron-glasses where temperature dependence was measured over a range of few
degrees around 4K it was found that while conductance increases
(exponentially) with temperature, relaxation remained constant, or even
becomes \textit{slower} \cite{24} with it. This work demonstrates the
distinction between conductance and relaxation in a different way; adding
carriers to the system by photo-excitation boosts one while impeding the
other. It is also important to note that the enhanced conductance makes it
unlikely that the dynamics slowdown in the PPC state is due to increased
disorder. While it is the interplay between interaction and disorder that
induces glassy behavior, it appears that the inter-particle interaction
(naturally being affected by a change in the carrier-concentration) is the
more dominant factor in slowing down relaxation rates.

The mechanism responsible for the detrimental effect of carrier-concentration
on relaxation may be either specific to the way $\Delta$n is generated [or
that $\Delta$n=$\Delta$n(t)], or it is of a generic nature. These are
discussed in turn.

Relaxation slowdown results from introduction of constraints on spatial
re-arrangement of charges. A mechanism specific to the PPC-state might result
from the local displacement of the ions from which charge was photo-generated
\cite{25}. It is not clear however that these defects could be more effective
in hindering charge organization than other defects so abundant in these
materials such as grain-boundaries, vacancies, etc. which are unaffected in
the dark-state$\rightarrow$PPC-state transition \cite{16}. Indeed, there is no
hint that these added defects play a role even in the conductance noise; the
amplitude of the 1/f power-spectrum is actually somewhat smaller in the PPC state.

Adding charge to a system naturally increases its conductance. In the hopping
regime this occurs by effectively reducing inter-site energy-differences.
However, it may also introduce another element that offsets this effect.
Doping of semiconductors for example also introduces scattering centers which
compromise the mobility. In the hopping regime added charge may decrease
tunneling probability by reducing wavefunctions overlap. Such a mechanism is
the Anderson orthogonality-catastrophe (AOC) \cite{26,24}. Localized-states in
Anderson insulators that exhibit long-lived MD are typically
\textit{multiply-}populated \cite{27}. These are essentially small disordered
metals which actually accentuates the AOC effect over the `clean limit'
\cite{28}. Tunneling probability between these mesoscopic elements, and the
presence of neighboring ones acting as bath, are reduced due to the AOC. The
reduction is more important the smaller is the "bare" tunneling probability,
slower transitions are made \textit{much} slower, fast ones are barely
affected \cite{24,28}. This \textit{non}-specific mechanism for the slowdown
does not depend on the method used to increase n, provided that the extra
charge is spread evenly throughout the sample volume (as in the process of
doping the sample or alloying it with another component that changes the
composition of the system). This requirement guarantees that the increase in
carrier-concentration includes the `dead-wood' regions. As noted above, the
slowest transition-rates involved in energy relaxation are due to the dynamics
in these regions.

The AOC scenario is supported by the observation that each and every Anderson
insulator with large carrier-concentration tested to date has exhibited
intrinsic EG effects (and indirectly, by the short relaxation times in
lightly-doped semiconductors \cite{15}). More experiments will be required to
elucidate these issues. In particular, using the PPC technique with materials
that are borderline in terms of their carrier-concentration and extending the
measurements to lower temperatures may produce useful information.

In summary, persistent-conductivity has been used to change the
carrier-concentration in the electron-glass phase of GeSb$_{\text{x}}%
$Te$_{\text{y}}$ films \cite{29}. This resulted in an enhanced conductance and
a substantial slowing-down of the relaxation dynamics relative to the
dark-state. This highlights the fundamental difference between conductance and
energy relaxation of electron-glasses and further illustrates the many-body
aspects of these phenomena. Neglecting many-body effects in treating the dc
conductivity of Anderson insulators may not lead to obvious discrepancies with
experiments. But using single-particle models to account for the long
relaxation times of the intrinsic electron-glasses is unlikely to yield much progress.

\begin{acknowledgments}
This research has been supported by a grant administered by the Israel Academy
for Sciences and Humanities. This research was supported in part by the
National Science Foundation under Grant No. NSF PHY11-25915
\end{acknowledgments}


\begin{thebibliography}{99}
\bibitem {29}As detailed in \cite{16} the stratagem employed here to modify
the carrier-concentration may in principle be used in other electron-glasses
provided they include local defects that sustain a barrier against recombination.
\end{thebibliography}
\end{document}